\documentclass[12pt,preprint]{aastex} 

\shorttitle{High energy emission} \shortauthors{Peer \& Waxman} 
 
\def\gsim{\;\rlap{\lower 2.5pt
 \hbox{$\sim$}}\raise 1.5pt\hbox{$>$}\;}
\def\lsim{\;\rlap{\lower 2.5pt
  \hbox{$\sim$}}\raise 1.5pt\hbox{$<$}\;}

\newcommand{\eV}{\rm{\, eV}} 
\newcommand{\KeV}{\rm{\, KeV}} 
\newcommand{\MeV}{\rm{\, MeV}}

\newcommand{\beq}{\begin{equation}} 
\newcommand{\eeq}{\end{equation}} 
\newcommand{\ba}{\begin{array}} 
\newcommand{\ea}{\end{array}} 
 
\begin{document}

\title{The high energy tail of $\gamma$-ray burst 941017:
 Comptonization of synchrotron self absorbed photons} 

\author{Asaf Pe'er\altaffilmark{1} and Eli Waxman\altaffilmark{1}} 
\altaffiltext{1}{Physics faculty, Weizmann Institute of Science, 
Rehovot 76100, Israel; asaf@wicc.weizmann.ac.il} 
 
\begin{abstract}
The recent detection of an unusually hard 
spectral component in GRB941017 extending to $\ge200$~MeV is hard to explain 
as a synchrotron emission from shock-accelerated electrons. It was argued to imply acceleration of protons to ultra-high energy. We show here  that the "high energy tail" can be explained as emission from shock-accelerated electrons in the early afterglow epoch, 
taking into account the effect of synchrotron self-absorption.
High energy observations set in this case stringent constraints on model parameters:
A lower limit to the total explosion energy  $E\gsim5 \times 10^{53}$~erg (assuming spherical symmetry); An upper limit to the density of gas surrounding the explosion,  $n\lsim10^{-2}(E/10^{54}{\rm erg}){\rm cm}^{-3}$; A lower limit to the expansion Lorentz factor $\Gamma_i\gsim 200$; and An upper limit to the fraction of thermal energy carried by the
magnetic field behind the shock driven into the surrounding medium, $\epsilon_{B,f} \le 10^{-4}$. Such constraints can not be inferred from keV--MeV data alone. The unusually low value of $\epsilon_{B,f}$ and the unusually high ratio $E/n$ may account for the rareness of GRB941017-type high energy tails. Tighter constraints on model parameters may be obtained in the future from optical and sub-TeV observations. 
\end{abstract}

 
\section{Introduction} 

In fireball models of GRBs \citep{Piran00,Meszaros02,Waxman03}
the energy released by an explosion is converted to kinetic energy 
of a thin baryonic shell expanding at an ultra-relativistic speed. 
The GRB is most likely produced by internal shocks within the
expanding shell. 
At a later stage, the shell impacts on surrounding gas, 
driving an ultra-relativistic shock into the ambient medium. 
This shock continuously heats fresh gas and accelerates relativistic
electrons, which produce synchrotron emission that account for 
the 
X-ray, optical and radio emission (the "afterglow")
following the GRB. 
The initial interaction of fireball ejecta with surrounding gas
produces a reverse shock which propagates into and decelerates 
the fireball ejecta \citep{MRP94}. 
As the reverse shock crosses the ejecta, it erases the memory 
of the initial conditions, and the expansion then approaches 
self-similarity \citep{BnM76}, 
where the expansion Lorentz factor decreases with radius, 
$\Gamma_{\rm BM}=(17E/16\pi n m_p c^2)^{1/2}r^{-3/2}$. 
Here $E$ is the total explosion energy (assuming spherical symmetry), 
$n$ is the number density of the ambient medium. 

The duration $T$ of the stage of transition to self-similar expansion, 
during which the reverse shock "lives", 
is comparable to the longer of the two time scales set by the initial
conditions \citep{Waxman03}: 
The (observed) GRB duration $T_{\rm GRB}$ and 
the (observed) time $T_{\Gamma}$ at which the 
self-similar Lorentz factor $\Gamma_{\rm BM}$ 
equals the original ejecta Lorentz factor $\Gamma_i$, 
$\Gamma_{\rm BM}(T_{\Gamma})=\Gamma_i$. 
Since the characteristic time over which radiation emitted by the
fireball at radius $r$ is observed by a distant observer is 
$\approx r/4\Gamma^2 c$ \citep{W97}, 
$T$ is determined by \citep{Waxman03}
\begin{equation}
T\approx 
\max\left[T_{\rm GRB}, 
10\left({E_{53} \over n_{-1}}\right)^{1/3}
\left({\Gamma_i\over300}\right)^{-8/3}{\rm\, s}\right],
\label{eq:T}
\end{equation}
where $E = 10^{53} E_{53}$~erg and $ n = 0.1 n_{-1} {\rm\, cm^{-3}}$. 
Note, that the duration is increased by a factor
 $1+z$ for a burst at redshift $z$.

Observations of GRB941017 show two distinct spectral components
\citep{Gonzalez}: 
A low energy component, with photon energies 
$\varepsilon_\gamma\lsim3$~MeV, 
and a high energy component, $\varepsilon_\gamma\gsim3$~MeV. 
The low energy component shows rapid variability, 
has a characteristic GRB spectrum peaking at 
$\varepsilon_\gamma\sim0.5$~MeV, and decays over $\sim100$~s. 
The high energy component has a very hard spectrum, 
number of photons per unit photon energy 
$dn_\gamma/d\varepsilon_\gamma\propto\varepsilon_\gamma^{-1}$, 
and persists over $\sim200$~s. 
The different temporal behavior suggests that the two components 
are produced in different regions of the expanding fireball. 
The characteristics of the low energy component suggest that 
it is produced by internal shocks, similar to other GRBs. 
The temporal behavior of the high energy component suggests 
that it is produced during the transition to self-similar expansion \citep{GnG}.

The hard, $dn_\gamma/d\varepsilon_\gamma\propto\varepsilon_\gamma^{-1}$, 
spectrum is difficult to account for in models where emission is 
dominated by shock accelerated electrons \citep{Gonzalez}. 
This has lead Gonz\'alez et al. to suggest that the high energy tail 
is due to electromagnetic cascades initiated by the interaction of 
photons with ultra-high energy shock-accelerated protons 
\citep{W95,Vietri,Dermer98,Totani98,WnB00}. We present here an alternative explanation:
Electrons accelerated in the forward shock
inverse-Compton scatter optical photons, emitted by the reverse shock
electrons, to create the observed spectra.
A key point, which allows to reproduce the observed hard spectrum, is the modification of the synchrotron spectrum by self-absorption in the reverse shock. We show that this effect allows to reproduce the observed hard, high energy tail also in the internal shock phase (see \S~\ref{sec:numeric}). However, we consider the latter explanation less likely, due to the weak time dependence of the high energy component.

\citet{GnG} have recently considered inverse-Compton emission from reverse shock electrons during the transition to self-similarity, as an explanation to the high energy tail of GRB941017. They have found that in order for such an explanation to be viable, the Lorentz factor associated with fireball expansion should be higher, $\Gamma\gtrsim10^{4}$, and the magnetic field in the fireball plasma should be much lower (well below equipartition), compared to values typically inferred from early afterglow observations \citep[see, e.g.,][]{ZKM03}. We show here that the high energy tail may be explained as emission from the forward shock electrons, with fireball plasma parameters which are typical to those inferred from GRB observations: $\Gamma\sim10^2$ and magnetic field close to equipartition (within the fireball plasma). In this scenario, the inferred density of plasma surrounding the fireball is lower than inferred for most other GRBs (the magnetic field strength in the shock driven by the fireball into the surrounding gas is poorly constrained by current observations). We consider the latter scenario more likely, since it requires a modification of the parameters of the environment external to the fireball, rather than modifications of the fireball physics. Our analysis further improves on that of \citet{GnG} in including the effects of self-absorption, and in carrying out detailed (numerical) calculations of the spectra, which are necessary given the inferred parameter range (see below). 

This paper is organized as follows. In \S~\ref{sec:plasma} we briefly discuss the dynamics of transition to self-similar expansion and the plasma conditions during the transition. In \S~\ref{sec:constraints} we analytically derive the constraints that should be satisfied by model parameters in order to allow an explanation of the observed high energy tail as inverse-Compton emission from forward shock electrons during
this phase.
In \S~\ref{sec:numeric} we present the results of detailed numerical calculations of the spectrum, which demonstrate that the observed spectrum may be reproduced when the constraints derived in \S~\ref{sec:constraints} are satisfied. Such calculations are necessary since the spectral shape near the inverse-Compton up-scattered self-absorption frequency is not well described by simple power-law approximations. Our conclusions are summarized in \S~\ref{sec:discussion}.

\section{The transition to self-similar expansion}
 \label{sec:plasma}
 
During the transition, plasma shocked by the reverse shock
expands with Lorentz factor close to that given 
by the self-similar solution, 
\beq
\Gamma \simeq 140 \, 
\left({E_{53} \over n_{-1}}\right)^{1/8}
T_2^{-3/8},
\eeq
where $T = 10^2 T_2 \rm{\, s}$. The unshocked fireball ejecta propagate with the original 
expansion Lorentz factor, $\Gamma_i > \Gamma$.
The forward shock which propagates into the ambient medium 
is highly relativistic, 
propagating with Lorentz factor $\approx\Gamma_{\rm BM}$, 
while the reverse shock is mildly relativistic, 
propagating into the ejecta with Lorentz factor 
$\approx1+\Gamma_i/\Gamma_{\rm BM}$. 

The radiation luminosity and spectrum are determined, 
for given $E$, $n$ and $\Gamma_i$, 
by the fraction $\epsilon_B$ ($\epsilon_e$) 
of shock thermal energy carried by magnetic field (relativistic
electrons), 
and by the shape of the electron distribution function, 
which is commonly assumed to be a power law of index 
$p\equiv-{\rm d}\ln n_e/{\rm d}\ln \gamma_e$ 
for electron Lorentz factors $\gamma_e$ exceeding $\gamma_{em}$. 
The values of the parameters $\epsilon_B$ and $\epsilon_e$ 
can not be determined from first principles, 
since the micro-physics of collisionless shocks which determines them 
is not fully understood. 
We therefore leave them as free parameters in the following analysis, 
and derive the constraints imposed on them by observations.
Since the  forward shock is highly relativistic,
while the reverse shock is only mildly relativistic,
we allow different values for $\epsilon_B$ and $\epsilon_e$ in the two shocks.
Thus, for example, 
$\gamma_{em,r} = \epsilon_{e,r} (m_p/m_e)(\Gamma_i/\Gamma_{\rm BM})(p-2)/(p-1)$
for the reverse shock, and 
$\gamma_{em,f} = \epsilon_{e,f} (m_p/m_e)\Gamma_{\rm BM}(p-2)/(p-1)$
for the forward shock (for $p>2$).

Synchrotron emission from electrons accelerated in the reverse shock 
typically peaks in the optical/IR band \citep{MR97,SP99}, 
\beq
\varepsilon_{\gamma,\rm RS}^{ob.} \approx 
\hbar \Gamma \gamma_{em,r}^2 {3 \over 2} {q B \over m_e c} = 
0.2 \, n_{-1}^{1/2} \Gamma_{i,2.5}^2 \epsilon_{e,r,-1}^{2} \epsilon_{B,r,-2}^{1/2} \eV.
\eeq
Here, $\Gamma_{i}=10^{2.5}\Gamma_{i,2.5}$, $\epsilon_{e,r}=10^{-1}\epsilon_{e,r,-1}$, and 
$\epsilon_{B,r}=10^{-2}\epsilon_{B,r,-2}$.
This energy is similar to the photon energy $\varepsilon_{\rm ssa}$ 
at which the optical depth for synchrotron self-absorption 
in the reverse shock equals unity,
\beq
\varepsilon_{\rm ssa}^{ob.} \approx 
0.8 \, E_{53}^{1/3} n_{-1}^{1/3} T_2^{2/3} \Gamma_{i,2.5}^{1/3} 
\epsilon_{e,r,-1}^{2/3} \epsilon_{B,r,-2}^{1/3} \eV.
\label{eq:ssa}
\eeq

Denoting by $\gamma_{c,r}$ the Lorentz factor of reverse shock electrons for which the 
synchrotron cooling time is comparable to the ejecta expansion time,
and assuming that electrons with  $\gamma > \gamma_{c,r}$ emit
nearly 100\% of their energy during the dynamical time,
the photon energy density is (for $p=3$)
\beq
u_{ph,syn,r} \approx 2 {\gamma_{em,r} \over \gamma_{c,r}} \epsilon_{e,r} u = 
2.5\times 10^{-2} \, E_{53}^{1/2}  n_{-1}^{3/2} T_2^{-1/2} \Gamma_{i,2.5} 
\epsilon_{e,r,-1}^{2} \epsilon_{B,r,-2} \rm{\, erg\, cm^{-3}},
\label{eq:u_ph}
\eeq
where the thermal energy density is $u = 4 \Gamma_{\rm BM}^2 n m_p c^2$.

\section{Constraints on model parameters}
\label{sec:constraints}

Electrons accelerated in the forward shock, with Lorentz factors 
$\gamma_{em,f}\sim \epsilon_{e,f}(m_p/m_e)\Gamma_{\rm BM}/2 \sim 10^{4.5}$, 
inverse-Compton scatter the optical photons, 
boosting their energy by $\sim\gamma_{em,f}^2$ to $\sim1$~GeV. 
The spectral shape of inverse-Compton scattered photons at observed energy range, 
$\varepsilon_\gamma<1$~GeV, is similar to the spectral shape of the reverse shock 
synchrotron spectrum at $\varepsilon_\gamma\lsim 1$~eV,  
as the energy of all photons is boosted by roughly the same factor
$\sim\gamma_{em,f}^2$. 
However, the spectral shape is expected to smoothen somewhat,
because Compton scattering is not exactly monochromatic.
The low energy synchrotron spectrum is 
affected by self-absorption (Eq. \ref{eq:ssa}).
The optical depth increases as photon energy decreases, leading to a thermal spectrum $dn_\gamma/d\varepsilon_\gamma\propto\varepsilon_\gamma^{1}$ at $\varepsilon_\gamma\ll\varepsilon_{\rm ssa}$. This spectrum is harder than observed. However, it becomes softer as $\varepsilon_\gamma$ approaches $\varepsilon_{\rm ssa}\approx\varepsilon_{\gamma,\rm RS}$, which may allow to produce a spectrum consistent with observations.

Thus, in order to explain the observed spectrum 
as due to inverse-Compton scattering of self-absorbed synchrotron
photons, the following conditions must be met:  
\begin{enumerate}

    \item The energy of synchrotron photons 
for which the reverse shock optical depth is unity, 
$\varepsilon_\gamma=\varepsilon_{\rm ssa}$, 
should be increased via inverse-Compton scattering by 
the lowest energy electrons in the forward shock, 
$\gamma_e=\gamma_{em,f}$, to $\approx 200 \MeV$, 
i.e. 
\beq
\gamma_{em,f}^2\varepsilon_{\rm ssa}\gtrsim200 \MeV.
\eeq

    \item The lowest energy electrons in the forward shock, 
$\gamma_e=\gamma_{em,f}$, should lose energy 
(via synchrotron and inverse-Compton scattering) 
on a time scale longer then $T$. 
If this condition is not met, cooling of the electrons would lead 
to a spectrum ${\rm d}\ln n_e/{\rm d}\ln \gamma_e\approx-2$ at 
$\gamma_e<\gamma_{em,f}$, 
which will produce an inverse-Compton spectrum 
$dn_\gamma/d\varepsilon_\gamma\propto\varepsilon_\gamma^{-3/2}$ 
below 1~GeV, softer than observed.
This condition can be written in the form
\beq
\ba{l}
\gamma_{c,f} \geq \gamma_{em,f}, \quad
\gamma_{IC,c,f} \geq \gamma_{em,f},
\ea
\eeq
where $\gamma_{IC,c,f}$ ($\gamma_{c,f}$)is the Lorentz factor of 
forward shocked electrons for which the cooling time due to inverse-Compton (synchrotron) emission is comparable to the expansion time.
The inverse-Compton cooling time is estimated via
$t_{cool,IC} = t_{cool,syn}(u_{B,f}/u_{ph,syn,r})$,
where $u_{B,f}=\epsilon_{B,f}u$ is the energy density in the magnetic field at the 
forward shock.

    \item Synchrotron emission from the forward shock electrons, 
as well as inverse-Compton emission from reverse shock electrons, 
should not exceed the observed flux at the 10~KeV to 1~MeV range.
These requirements can be met by requiring either (a) The flux to be lower than observed, or (b) The photons to be emitted below 10~KeV. 
For synchrotron emission from the forward shock this implies either
\beq
{L_{IC,f} \over L_{syn,f}} = {u_{ph,syn,r}\over u_{B,f}} > 10^3, \, {\rm or} \quad 
\varepsilon_{\gamma,\rm FS}^{ob.} \lesssim 1 \KeV . 
\eeq

For inverse-Compton emission from the reverse shock it implies either 
\beq
{ L_{IC,f} \over L_{IC,r}} = 
{\epsilon_{B,r} \over \epsilon_{e,r}}{\gamma_{c,r} \over \gamma_{em,r}} 
 \gamma_{em,f}^2 \tau_{T,f} > 10^3 , \, {\rm or} \quad 
\varepsilon_{\gamma, IC,\rm RS}^{ob.} = 
\varepsilon_{\gamma,\rm RS}^{ob.} \gamma_{em,r}^2 \lesssim 1 \KeV.
\eeq


\end{enumerate}

A lengthy, but straight forward, manipulation of the above constraints lead to the following constraints on model parameters\footnote{ A detailed description of the derivation may be found 
at http://www.weizmann.ac.il/$\sim$asaf/ }:
\beq
\ba{l r c l }
a. & E_{53} & \geq & 4 \, T_2^{3/2}, \\
b. & \left({E_{53} \over {n_{-1}}} \right) &\geq &50 \, T_2^4, \\
c. & \epsilon_{e,r,-1} &\geq &1.5 E_{53}^{-1} \, T_2^{5/2}, \\
d. & \epsilon_{B,r,-2} &\lesssim& 15 \, T_2^{37/8}, \\
e. & \epsilon_{e,f} &\simeq& 0.33,  \\
f. & \epsilon_{B,f,-2}& \leq& \left( 2.5 \times 10^{-4}, 5 \times 10^{-4}\,T_2^{3/4}, 5 \times 10^{-3}\, T_2^{1/2}, 1.2 \times 10^{-2} \, T_2^{5/4} \right),  \\
g. & \Gamma_{i,2.5}& \geq& 0.76 \,  T_2^{1/8}.
\ea
 \label{eq:constraints}
\eeq
It is sufficient to meet 1 of the 4 conditions (f). It should be pointed out here that the redshift of the source of GRB941017 is considered a free parameter in the above analysis. A particular choice of model parameters implies a particular choice of redshift given the measured flux, as illustrated in few examples in figure~\ref{fig1}.

\section{Numerical results}
\label{sec:numeric}

In order to obtain an accurate description of the spectrum, 
we have carried detailed numerical calculations of the emission 
of radiation from a fireball in transition to self-similar behavior. 
Our numerical model contains full description 
of cyclo-synchrotron emission, synchrotron self absorption,
inverse- as well as direct Compton scattering, pair production and 
annihilation and cascade processes occurring at high energies. Full description of the model appears in \citet{PW03}.
In the present version, photons emitted after acceleration of electrons at the 
reverse shock, serve as seed photons for physical processes occurring
after acceleration of electrons in the forward shock.

Figure~\ref{fig1} presents three examples of the results of our
calculations for three different sets of parameter values, consistent with the constraints derived in \S~\ref{sec:constraints}. It demonstrates that the process described above may indeed 
account for the observed hard spectrum of the high energy component. 

An alternative explanation of the spectrum, emission from internal collisions within the expanding fireball, 
is also presented. For this scenario, we assumed $\epsilon_B$ and $\epsilon_e$ to be similar in the two shock waves created as two plasma shells collide. The observed spectrum may be reproduced adopting rapid variability, $\Delta t_{var} \approx 10^{-5}$~s, and very low value of the equipartition fraction, $\epsilon_B \approx 10^{-7}$. Here too, the hard high energy tail is due to inverse-Compton scattering of self-absorbed synchrotron spectrum. Accounting for $\sim1$~MeV GRB emission, like the variable low energy component observed at the first $\sim 100$~s in GRB941017, typically requires in the internal shock scenario equipartition fraction close to unity. This, and the temporal characteristics of the high energy component, suggest that the high energy tail is produced during the transition to self-similar expansion, rather than in internal shocks.

The following point should be noted here. \citet{Gonzalez} present spectra averaged over 5 distinct time intervals. The flux carried by the high energy tail shows variations between different time intervals, while large uncertainties in the determination of the spectral index in individual time intervals make it difficult to determine
whether the spectral index is time dependent. In principle, time dependence of the high energy flux and spectrum provides additional constraints on the model. Such variations would arise in the model due to the expansion of the plasma during the transition phase, and due to deviations from homogeneity in the fireball shell as well as in the surrounding medium. Thus, the observed time dependence may allow to constrain the distributions of density and Lorentz factor within the fireball ejecta, and the distribution of ambient medium density. Given the relatively large uncertainties in the determination of the time dependence of the flux and of the spectral index, we find that such analysis is not warranted by the current data.

\section{Discussion}  
\label{sec:discussion}

\begin{figure}
\plotone{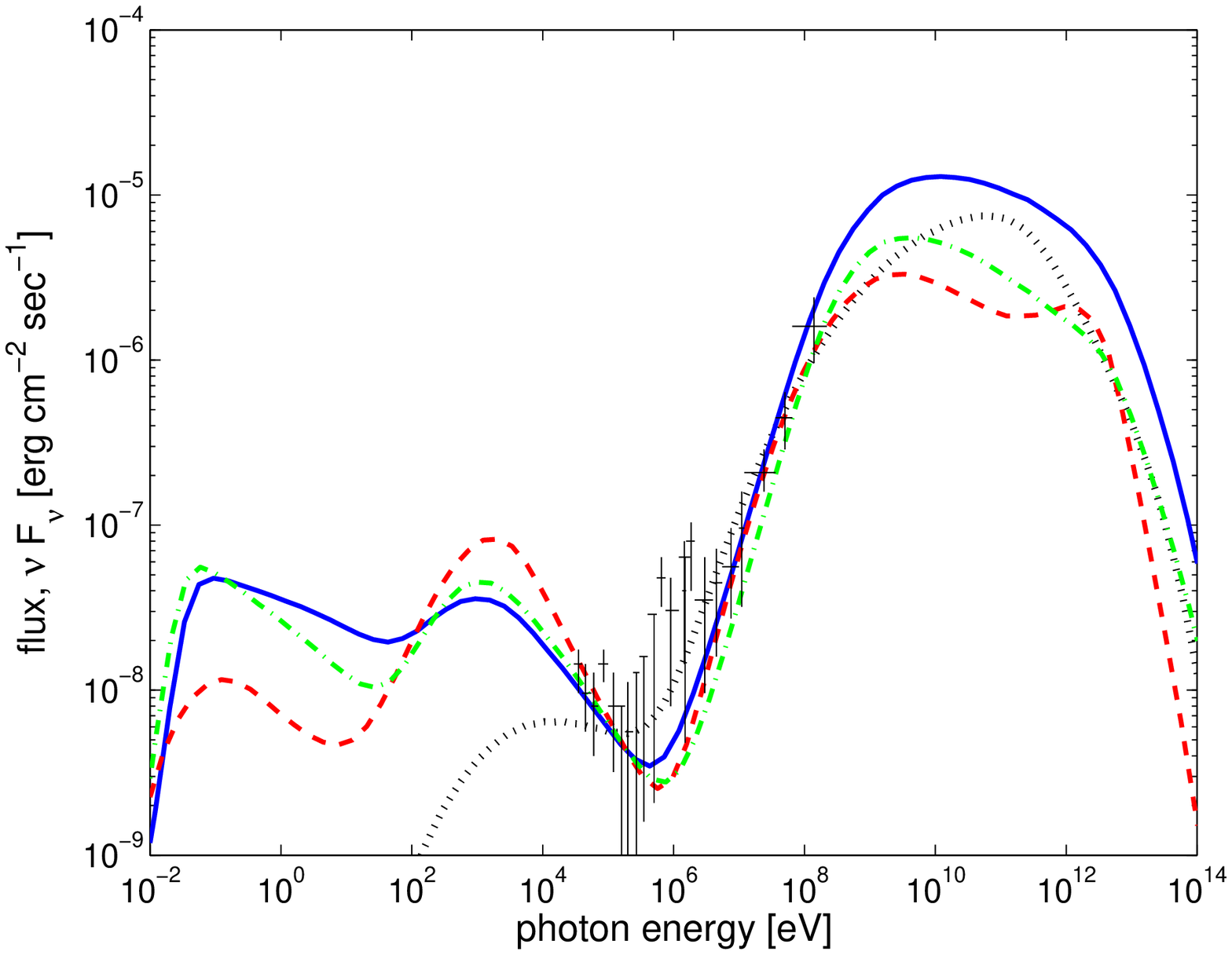} 
\caption{Points with error bars are the observed spectrum of GRB941017,
averaged over a time interval of 100~s to 200~s 
following the burst trigger \citep{Gonzalez}. 
Detailed calculations are presented for three different sets of model 
parameters, 
which are consistent with the constraints of Eq.~\ref{eq:constraints}: 
$\{E=3\times10^{54}{\rm\, erg},n=0.1{\rm\, cm}^{-3},\Gamma_i=300, 
\epsilon_{B,r} = 10^{-1}, \epsilon_{B,f} = 10^{-6}, z=0.1\}$ 
(solid),
$\{E=10^{55}{\rm\, erg},n=0.1{\rm\, cm}^{-3},\Gamma_i=200, 
\epsilon_{B,r} = 10^{-3}, \epsilon_{B,f} = 10^{-5}, z=0.25\}$ 
(dashed) and 
$\{E=10^{54}{\rm\, erg},n=0.03{\rm\, cm}^{-3},\Gamma_i=220, 
\epsilon_{B,r} = 0.2, \epsilon_{B,f} = 3 \times 10^{-6}, z=0.06\}$ 
(dash-dotted).
The dotted line presents an alternative explanation: rapid variability in prompt emission, with
model parameters: $L=3\times10^{52}{\rm\, erg \, s^{-1}},\Delta t_{var}=10^{-5} {\rm\, s},
\Gamma_i=1500, \epsilon_{B,r} = \epsilon_{B,f} = 10^{-7}$, $z=0.15$.
}
\label{fig1}
\end{figure}

We have shown that the high energy spectral component of GRB941017 may be explained as inverse-Compton emission from electrons accelerated by the shock driven by the fireball into its surrounding medium during
the transition to self-similar expansion (Figure~\ref{fig1}). The high energy tail may also be explained as inverse-Compton emission from electrons accelerated in internal shocks (Figure~\ref{fig1}), and in the reverse shock during transition to self-similar expansion \citep{GnG}. As explained in the introduction, we consider the latter explanations less likely, since they require modifications of the fireball physics (in particular, magnetic field well below equipartition in the fireball plasma), and since the temporal behavior is hard to account for in the internal shock scenario.

We have shown that the high energy data, when interpreted as emission from forward shock electrons, allow to put stringent constraints (Eq.~\ref{eq:constraints}) on fireball parameters, such as the total (isotropic equivalent) energy $E$, initial Lorenz factor $\Gamma_i$ and ambient medium density $n$. Such constraints can not be inferred from keV---MeV observations alone. The inferred values of $E$ and $\Gamma_i$ are similar to those typical to cosmological GRBs, while the inferred constraint $E/n\gtrsim10^{56}{\rm erg\,cm}^3$ implies a ratio $E/n$ which is higher than typically obtained, $E/n\sim10^{54}{\rm erg\,cm}^3$ \citep{Bloom03}. The inferred values of $\epsilon_{e,f}$, $\epsilon_{B,r}$ and $\epsilon_{e,r}$ are consistent with those inferred from other GRB and afterglow observations. The value of $\epsilon_{B,f}$ is usually less well constrained by
observations, and is required to be well below equipartition in our case. This, and the large ratio of $E/n$ may account for the rareness of GRB941017-type high energy tails.

The unknown redshift of the source of GRB941017 is treated in our analysis as a free parameter: A choice of model parameters implies a choice of redshift
(see figure~\ref{fig1}). Values of $E$ close to the lower limit, $\simeq10^{54}$~erg, imply
low redshift, $z\sim0.1$. This is a direct consequence of the fact that the gamma-ray fluence of GRB941017 is unusually high, rather than of the fact that its high energy spectrum is unusually hard. The fluence, $6.5\times10^{-4}{\rm erg/cm^2}$, implies an (isotropic equivalent) gamma-ray energy release of $\sim10^{55}$~erg for $z=1$.

Figure~1 demonstrates that different scenarios accounting for the high energy component of GRB941017, as well as different model parameters in a given scenario, lead to different model predictions for the fluxes at optical, X-ray and sub-TeV energy bands. 
The predicted fluxes
are well within the detection capabilities of the SWIFT (in optical) and GLAST (at 10--100GeV) satellites, and of sub-TeV ground based Cerenkov telescopes 
(e.g. HESS\footnote{http://www.mpi-hd.mpg.de/hfm/HESS/HESS.html}
, MAGIC \footnote{http://hegra1.mppmu.mpg.de/MAGICWeb}
, MILAGRO  \citep{McEnery03}, VERITAS \citep{Weekes02}). Therefore, optical and sub-TeV observations on minute time scale will allow to put more stringent constraints on explosion parameters than those given in Eq.~\ref{eq:constraints}. 

\acknowledgements
We thank J. Granot \& D. Guetta for helpful discussions.
This work was supported in part by a Minerva grant and by a grant from the Rosa and Emilio Segr\'e fund.

\end{document}